\documentclass[12pt]{iopart}
\usepackage{epsfig}
\begin{document}
\title{Is the Radion a Chameleon?}
\author{Ph.~Brax}
\address{Service de Physique Th{\'e}orique, CEA--Saclay, Gif/Yvette cedex, 
France F--91191}
\author{C.~van de Bruck}
\address{Astro--Particle Theory and Cosmology Group, 
Department of Applied Mathematics\\
University of Sheffield, Sheffield S3 7RH, UK}
\author{A.--C.~Davis}
\address{DAMTP, Centre for Mathematical Sciences, Cambridge University\\
Wilberforce Road, Cambridge CB3 0WA, UK}
\begin{abstract}
The chameleon mechanism is a way to give an effective mass to a light 
scalar field via field self-interaction and interaction with matter. 
We study this mechanism in models in which the couplings are field 
dependent and find that the properties are very different from the 
case of constant couplings. The consequences of a runaway 
potential for the radion field in brane world scenarios and whether the radion 
can play the r{\^o}le of dark energy is investigated. The cosmological 
evolution during the inflationary epoch, the radiation and the 
matter dominated epochs are discussed as is the compatibility of 
the radion field with local tests of gravity.
\end{abstract}

\section{Introduction}
Current cosmological observations suggest that the expansion of
the universe is accelerating instead of slowing down (see e.g. 
\cite{supernovae} for recent results). The type of
matter responsible for the accelerated expansion has been dubbed
{\it dark energy}. Not much is known about the nature this new form
of energy but different candidates have been suggested such as a
cosmological constant or a slowly rolling scalar field (also called
quintessence--field or cosmon--field) (see \cite{quintessence1} and 
\cite{quintessence2} for early work and e.g. \cite{dereview1,dereview2} for 
recent reviews). A challenge for particle physics is to
identify candidates with the right properties. For example, the
coupling of any new matter type to ordinary matter cannot be too
large, otherwise dark energy would have been detected in the form
of an additional force acting on bodies in laboratory experiments.

If current observations are confirmed by future experiments, it is
highly probable that an explanation for dark energy will involve
new physics beyond the standard models of particle physics and
cosmology. One possible type of new physics beyond the standard
model is the idea of extra dimensions. Among the models with
higher dimensions, brane worlds have attracted a lot of attention
recently. For recent reviews see \cite{branereview1}--\cite{branereview4}. It has
been speculated that the current accelerated expansion is due to
the effects of extra spatial dimensions \cite{de-edim1,de-edim2,de-edim3,
de-edim4,de-edim5,de-edim6}. Such models
can be tested (or constrained) with the observation of the CMB
anisotropies, see e.g. \cite{cmb1} and \cite{cmb2}. Higher dimensional theories provide a
plethora of scalar fields which usually couple to matter. String
theory, for example, predicts the existence of the dilaton field,
which couples universally to all matter forms. Another example is
the brane world theory, according to which our universe is a
hypersurface embedded in a higher dimensional space. There, scalar
fields in the bulk and/or the size of the extra dimensional
space-time manifest themselves as scalar fields in the low--energy
description of the system. Usually, all these fields couple to all
matter forms confined on the brane. It is so far not known if
these fields can play the r{\^o}le of dark energy and, at the same
time, remain hidden from experiments on earth. Several proposals
have been made in the past, in which the quintessence field
couples to baryons and/or dark matter (see e.g. 
\cite{coupleddarkenergy1}--\cite{coupleddarkenergy4} and references therein).
Cosmological considerations, such as nucleosynthesis, the anisotropy 
spectrum of the cosmic microwave background radiation, structure 
formation, etc., constrain the coupling of the quintessence field 
to matter (baryonic or non--baryonic).

In this paper we investigate a system of two branes and bulk
scalar field as a specific setup of a higher--dimensional theory. 
We discuss under which conditions the radion, i.e.
the interbrane distance, can be a candidate for dark energy and if,
at the same time, the theory is compatible with local experiments. We are 
going to assume that a potential for the radion field has been 
generated, for example by non--perturbative effects. Therefore, 
in our model the dynamics of the radion is governed by scalar field self-interactions
and interaction with surrounding matter. Such fields, in which 
the effective potential is matter (environment) dependent, have recently been dubbed {\it
chameleon fields} and arise usually in scalar--tensor theories with a 
potential for the scalar field \cite{chameleon}. We
show that for scalar--tensor theories with an essentially linear 
coupling function, one can evade local tests thanks to the 
existence of a thin--shell regime: the chameleon field is 
basically constant  inside a big, spherical body and varies only 
inside a thin--shell close to the outer edge of the body. This 
implies that the fifth force outside the body is a small 
correction to Newton's law. 

There are two regimes for the radion in the brane model we consider. 
At small distances (i.e. distances smaller than the bulk curvature scale), 
the radion possesses the thin--shell property while at large distance (distances 
larger than the bulk curvature scale) no thin-shell regime exists.
The thin--shell case is expounded in \cite{chameleon,chameleoncosmology}. 
Here we focus on the case of thick--shells at large distances 
between the branes.

The paper is organized as follows: In Section 2 we describe
briefly the chameleon properties. In particular, thin and thick
shells are discussed. In Section 3 we focus on the radion and show
that both possibilities are realised at either small or large
distances. In Section 4 we investigate first the inverse
power--law potential in detail. After discussing the cosmological
evolution, we elaborate on local experiments and constraints.
These constraints are stringent. We then generalise these results
to a class of potentials where, at large distances 
between the branes, the radion can both be a quintessence field and comply
with gravity experiments. In Section 5 we discuss a non--standard potential 
and the consequences for the evolution of the radion. We present our conclusion \
in Section 6.

\section{Chameleon Fields}
In this Section, we discuss the properties of chameleon fields.
Because these fields appear naturally in scalar--tensor theories
of gravity in which the scalar field has a potential, we begin 
with a discussion of these theories.
\subsection{Scalar--tensor theories}
We consider theories where a scalar field $\phi$ couples both to
gravity and matter, generating a potential fifth force. The
Lagrangian of such scalar--tensor theories reads
\begin{equation}
S = \frac{1}{2\kappa_4^2}\int d^4 x\sqrt{-g} (R- (\partial \phi)^2
-2\kappa_4^2 V(\phi))
\end{equation}
Notice that $\phi$ is dimension--less. Redefining $\phi \to
\kappa_4^{-1}  \phi$ leads to a dimension--one field. 
Here, $\kappa_4= 1/ m_{\rm Pl}$, where $m_{pl}$ is the reduced 
Planck mass. Matter couples to both gravity and the scalar field
according to
\begin{equation}
S_m(\psi, A^2(\phi)g_{\mu\nu}),
\end{equation}
where $\psi$ is a matter field and $A$ is an arbitrary function of
$\phi$. The Klein--Gordon equation for the scalar field involves a
coupling to the trace of the matter energy--momentum tensor $T$
\begin{equation}
D^{\mu}D_{\mu} \phi =\kappa_4^2 \frac{\partial V(\phi)}{\partial
\phi}-\kappa_4^2\alpha_\phi T,
\end{equation}
where
\begin{equation}
\alpha_\phi= \frac{\partial \ln A}{\partial \phi}.
\end{equation}
The energy--momentum tensor in the Einstein frame fulfills
\begin{equation}
D_\mu T^\mu_{_\nu}= \alpha_\phi \left(D_\nu \phi \right) T.
\end{equation}
In a Friedmann--Robertson--Walker Universe these equation simplify to
\begin{equation}
\ddot \phi + 3H \dot \phi = -\kappa_4^2\frac{\partial
V(\phi)}{\partial \phi} - \kappa_4^2 (1-3w_m) \rho\frac{\partial A}{\partial\phi},
\end{equation}
together with
\begin{equation}
\dot \rho +3H \rho =\alpha_\phi  \dot \phi \rho.
\end{equation}
Redefining
\begin{equation}
\rho=A(\phi) \rho_m,
\end{equation}
one can easily verify that
\begin{equation}
\rho_m=\frac{\rho_0}{a^{3(1+w_m)}}
\end{equation}
is the usual energy density.

The Klein--Gordon equation can be written in terms of an effective potential
\begin{equation}
\ddot \phi + 3H \dot \phi = -\kappa_4^2\frac{\partial
V_{\rm eff}(\phi)}{\partial \phi},
\end{equation}
where
\begin{equation}
V_{\rm eff}(\phi)=V(\phi) +\rho_m(1-3w_m)  A(\phi).
\end{equation}
The effective potential depends on the environment through
$\rho_m$.  In particular, we will assume that $V(\phi)$ is a runaway
potential and for the models we consider $A(\phi)$ increases with $\phi$.
The potential has a minimum for
\begin{equation}\label{minimum}
\frac{V'(\phi_{\rm min})}{A'(\phi_{\rm min})}=-\rho_m (1-3w_m).
\end{equation}
Such a field has been called a chameleon field. The field, when
sitting at the minimum, becomes massive with mass $(\kappa_4^2
V_{\rm eff}^{''})$
\begin{equation}
m^2\vert_{\phi_{min}}=-\kappa_4^2 \alpha_\phi(1-3w_m) A \frac{d\rho}{d\phi(\rho)}\vert_{\phi_{min}},
\end{equation}
evaluated at the minimum. In this equation, we have made use of the
fact that, at the minimum, $\phi$ is a function of $\rho$, as seen
from equation (\ref{minimum}).

Dimensionally the mass is proportional to $\kappa_4^2
\rho_m\approx H^2$ on cosmological scales, i.e.  the field at the
minimum has a very small mass. Naively, one would expect that
in that case solar system experiments  would lead to very strong
constraints on the models. However, because the effective mass
depends on the energy density of the environment, this is not
necessarily true.

\subsection{Solar Tests}
In the following, we impose constraints springing from local
experiments, i.e. solar system experiments. The field $\phi$ acts
on all types of matter and, in the Einstein frame, there is a new
force associated with the scalar field. Therefore, massive
particles no longer follow geodesics with respect to the Einstein
frame metric.  The equations of motion in the Einstein frame are
\begin{equation}
\ddot x^\mu + \Gamma^{\mu}_{\nu\lambda}\dot x^\nu \dot x^\lambda
 + \alpha_\phi \frac{\partial \phi}{\partial x_\mu} = 0
\end{equation}
In this equation, $x^a$ represents the four--vector describing the
worldline of a test particle moving in the metric $g_{\mu\nu}$,
$\Gamma^{\mu}_{\nu\lambda}$ represents the Christoffel symbols for the
metric $g_{\mu\nu}$. The last term is a new contribution,
representing the force from the scalar field, which is given by
\begin{equation}\label{fifthforce}
F_{\phi} = - m \alpha_\phi\frac{\partial \phi}{\partial x_\mu},
\end{equation}
where $m$ is the mass of the test particle. The force $F_{\phi}$
cannot be too large, otherwise experiments would have already
detected it. What is the profile $\phi(r)$ for an object like the
earth? Modelling the earth as a spherical object with constant
density $\rho_{\bullet}$,  the equation for $\phi$ (i.e. the
Klein--Gordon equation), reads for $r\leq R_{\bullet}$
\begin{equation}
\phi^{''} + \frac{2}{r}\phi' =\kappa_4^2 \frac{\partial
V_{\rm eff}}{\partial \phi}, \label {in}
\end{equation}
whereas for $r>R_{\bullet}$ we assume that the field is close to
its minimum $\phi_\infty$ determined cosmologically (we assume
that space is empty outside the body)
\begin{equation}
\phi^{''} + \frac{2}{r}\phi' =  m_\infty^2 (\phi
-\phi_\infty)
\end{equation}
For the function $A(\phi)$, we will focus on the form
\begin{equation}\label{coupling}
A=1+\beta \frac{\phi^{m+1}}{m+1},
\end{equation}
 In this paper, we will consider the case in which $\beta$ is of
order one. As we will show when  $m=0$ a thin--shell regime
exists, whereas for $m=1$  there are no thin shells even for
massive bodies. The case $m\ge 1$ is pathological as it leads to
no thin-shell and an infinite repulsive force in the core of the
spherical body. Putting $\phi=r^{-1}\psi$ we get
\begin{equation}
\psi^{''} =\beta \kappa_4^2 \rho_\bullet \frac{\psi^m}{r^{m-1}}.
\end{equation}
Let us discuss the case $m=0$ first.
\subsection{The thin--shell regime}
We will solve the equations of motion inside and outside the
spherical body. We will assume that the minimum outside the body
$\phi_\infty$ is greater that the minimum inside $\phi_b$. This is
the case for simple potentials like an inverse power law.

First of all, we impose that $\phi'=0$ at $r=0$. Interpreting the
equations of motion as classical mechanics in the  effective
potential $-V_{eff}(\phi)$, this corresponds to no velocity at the
origin. Moreover we fix the value of the field there to be  near
the minimum $\phi_b$ of the effective potential inside the body.
The solution can be approximated in three different regions. First
for $r\le R_s$ the field is constant $\phi=\phi_b$. In a shell
$R_s\le r\le R_\bullet$ the field is increasing in order to match
the value $\phi_\infty >\phi_b$. It reads
\begin{equation}
\phi=\phi_b - \frac{3\beta}{8\pi} \frac{\kappa_4^2M_s}{R_s} + \frac{\beta
\kappa_4^2 \rho_\bullet}{6}r^2 +\frac{\beta
\kappa_4^2}{4\pi}\frac{ M_s}{r}.
\end{equation}
 Outside the body for $r>R_\bullet$ the
field behaves like a Yukawa potential
\begin{equation}
\phi=\phi_\infty +\frac{\beta \kappa_4^2}{4\pi}
\frac{M_s-M_\bullet}{r} e^{-\kappa_4 m_\infty (r-R_\bullet)}.
\end{equation}
In the vicinity of the body and using the fact that
$m_{\infty}\approx H_0$ we can neglect the exponential profile.
Hence the outside the body the force is a correction to Newton's
law. Matching at $R_\bullet$ leads to
\begin{equation}
\frac{R_\bullet-R_s}{R_\bullet} = \frac{1}{6\beta} \frac{\phi_\infty - \phi_b}{\Phi_0},
\end{equation}
where $\Phi_0 =\kappa_4^2  M_\bullet/8\pi R_\bullet$ is the
gravitational potential on the surface of the body. Because $\phi$
can be viewed as a potential for a new, fifth force (see eq.
(\ref{fifthforce})), the above equation says that if the ratio
between the difference of the field at infinity and inside the
body and the gravitational potential is small, $R_s$ is of order
$R_\bullet$. This regime has been called the {\it thin--shell
regime}. It is realised when
\begin{equation}
(\phi_\infty -\phi_b) \ll \beta \Phi_0.
\end{equation}
The new force leads to a modification to Newton's law outside the body
\begin{equation}
F=(1+\theta)F_{Newton},
\end{equation}
where $F_{Newton}= -mM_\bullet G_N/r^2$ and
\begin{equation}
\theta=\frac{9G_N(\phi_{\infty}-\phi_b)}{R_\bullet^2}.
\end{equation}
which is suppressed
\begin{equation}
\vert \theta \vert \ll  \frac{\rho_\bullet}{m_{\rm pl}^4} \ll 1.
\end{equation}
 Despite being extremely light, the chameleon field does not
generate a measurable fifth force. This results springs from the
smallness of $R_\bullet -R_s$, i.e. the thin--shell regime.

To conclude, in the case $m=0$ we have found that, like in the
original chameleon model, the thin--shell regime may exist. We will see that in the
case $m=1$, the thin--shell regime does not exist.

\subsection{The thick--shell regime}

Let us repeat the same analysis for $m=1$. In the range $R_s\le r\le  R_\bullet$
we find that $\phi$ is a combination of $\frac{\sinh
(r\sqrt{\kappa_4^2\rho_\bullet})}{r}$ and $\frac{\cosh
(r\sqrt{\kappa_4^2\rho_\bullet})}{r}$. Matching at $R_s$ leads to
\begin{equation}
R_s=\sqrt{\frac{2}{\kappa_4^2 \rho_\bullet}},
\end{equation}
implying that $R_s\gg R_\bullet$. Thus, there is no thin--shell regime,
even inside big, heavy bodies.

Demanding that the first derivative vanishes at the origin, the solution inside the
body can be found to be
\begin{equation}
\phi_{I} = \tilde{A}\frac{\sinh(r/R_r)}{r},
\end{equation}
where $R_r^2 = m_{pl}^2 / \beta \rho_{\bullet}$. Outside the body,
the solution reads
\begin{equation}
\phi_{E} = {\cal C} + \frac{\cal D}{r}.
\end{equation}
Matching both solutions and their first derivative at
$r=R_{\bullet}$ and noting that $R_{\bullet}\ll R_r$ and $\phi
\rightarrow \phi_{\infty}$ as $r \rightarrow \infty$ one obtains
\begin{eqnarray}
\tilde{A} &=& -3{\cal D}\frac{R_r^3}{R_{\bullet}^3}, \\
{\cal C} &=& \phi_{\infty}, \\
{\cal D} &=& \frac{\phi_\infty R_{\bullet}}{1 +
3\frac{R_r^2}{R_{\bullet}^2}},
\end{eqnarray}
where $\phi_{\infty}$ is the cosmological value of the field now,
i.e. the value of the field far away from the body.

Thus, away from the body, the profile is given by
\begin{equation}
\phi(r) = \phi_{\infty}-\frac{\beta
\phi_\infty}{4\pi}\frac{M_{\bullet}}{m_{pl}^2} \frac{1}{r},
\end{equation}
leading to a force
\begin{equation}
F_{\phi}=-\frac{\beta^2\phi^2_\infty}{4\pi}\frac{mM_{\bullet}}{m_{pl}^2 r^2},
\end{equation}
which is nothing but a correction to Newton's law with
\begin{equation}\label{radioncoupling}
\theta=2\beta^2 \phi_\infty^2.
\end{equation}
Thus, in the case with $m = 1$, the effective coupling 
between scalar field and matter is not specified by $\beta$ alone, {\it it is 
also determined by the cosmological value of the field itself.}
Experiments constrain $\theta \le 10^{-5}$, see \cite{st-test1} and 
\cite{st-test2}. This implies that the value
of the field at infinity, determined by the cosmological
dynamics, must be extremely small.

\subsection{The thick--shell catastrophe}
We will briefly discuss the case $m > 1$. The solution inside the
body reads
\begin{equation}
\phi= F_0 r^{2/(1-m)}
\end{equation}
where $F_0^{m-1}=\frac{2(3-m)}{(1-m)^2\beta \kappa_4^2
\rho_\bullet}$.
This solution is only valid when $m$ is even, the $(m-1)$--th root
  taking alternative signs.
Notice first that this solution cannot match a constant between $R_s$
  and $R_\bullet$. Hence there is no thin--shell and the above
  solution is valid all the way to the origin.
At the origin the field diverges
and the force behaves like
\begin{equation}
\frac{F}{m}=O\left(\frac{1}{r}\right)
\end{equation}
It is repulsive and infinite at the origin implying that the compact
spherical object would not exist.

In conclusion, only $m=0,1$ are sustainable cases. Whereas the case $m=0$ 
allows for a thin--shell regime, the case $m=1$ does not have this propery. 
The case $m>1$ is pathological and does not lead to physical acceptable 
results. 

\section{The Radion}
Let us now
turn our attention to a scalar--tensor theory motivated from the
brane world idea. In what follows, we will in particular focus on
the radion field, i.e. the interbrane distance. Let us first
concentrate on two-brane models with a bulk scalar field (for details, see e.g. 
see \cite{moduli0}, \cite{moduli1} and \cite{moduli}). At low
energy the model is described by a tensor--scalar field theory
with two massless scalar fields, ${\cal R}$ and ${\cal S}$, related to the scale
factor in the bulk $a(z)$
\begin{equation}
\tanh \frac{\cal R}{\sqrt 6} =\frac{a^{1+2\delta^2}(z_2)}{a^{1+2\delta^2}(z_1)}
\end{equation}
where $z_{1,2}$ are the brane coordinates and
\begin{equation}
{\cal S}=\frac{1}{2} \ln (a^{2+4\delta^2}(z_2)-a^{2+4\delta^2}(z_1))
\end{equation}
For $\delta=0$, the second field ${\cal S}$ decouples and we obtain the
Randall-Sundrum model \cite{RS} with a single field
\begin{equation}
d=-l\ln\left(\tanh \frac{\cal R}{\sqrt 6}\right)
\end{equation}
where $d$ is the interbrane distance, i.e. the radion, and $l=1/k$ measures the curvature scale of the bulk space-- time. However, even if
$\delta \ne 0$  we will still call ${\cal R}$ the radion field.

In the bulk, the scale factor behaves like
\begin{equation}
a(z)=(1-4k\delta^2 z)^{1/4\delta^2}
\end{equation}
showing that there is a would--be singularity when $a=0$ beyond
the second brane. When coupled to matter, the cosmological
evolution drives ${\cal R}$ to zero, i.e. the second brane hits the
would--be singularity \cite{moduli, moduli0}.

This situation can be avoided by introducing a potential $V(r)$
shielding ${\cal R} = 0$. Such a run--away potential can be chosen to be like
\begin{equation}\label{poti}
V({\cal R})=\frac{\Lambda^4}{({\cal R}-{\cal R}_*)^\alpha}.
\end{equation}
where ${\cal R}\ge {\cal R}_*$ is the allowed range.  As we assume that ${\cal S}$ 
is still massless, solar system experiments imply that $\delta \ll
1$, i.e. the Randall--Sundrum limit. In that case the coupling of the field ${\cal R}$
to matter is well specified by the function \cite{moduli}
\begin{equation}
A({\cal R})=\cosh\left(\frac{\cal R}{\sqrt{6}}\right).
\end{equation}
In the Randall-Sundrum case, there are two relevant regimes. When
${\cal R}\gg 1$, the distance between the branes is small $d\ll l$ (i.e. 
the distance is smaller than the bulk curvature length scale).
Changing ${\cal R}\to \tilde {\cal R}={\cal R}-{\cal R}_*$, the coupling becomes then
\begin{equation}
A({\cal R}) =\frac{1}{2} e^{\frac{{\cal R}+{\cal R}_*}{\sqrt 6}}
\end{equation}
In that case, we retrieve the original example of a chameleon
field corresponding for ${\cal R}_*\gg 1$ and $\tilde {\cal R} \ll 1$ to a linear coupling with $m=0$. 
Hence the radion possesses then a thin--shell. This situation has been studied at length 
in \cite{chameleon}. Now there is another interesting regime when ${\cal R} \ll 1$ and $d\gg l$ (i.e. 
interbrane distances larger than the bulk curvature scale), where we
can approximate
\begin{equation}
A \approx 1+\frac{{\cal R}^2}{12},
\end{equation}
and we have taken ${\cal R}_*=0$.  It corresponds to
\begin{equation}
m=1 \mbox{ } \hspace{0.2cm} {\rm and} \hspace{0.2cm} \beta = \frac{1}{6}
\end{equation}
in equation (\ref{coupling}).

Thus, if $\cal{R}$ is small enough with a runaway potential of the form (\ref{poti}), we end up with
a theory in which the thin--shell regime does not exist even for heavy bodies. Cosmology has
to ensure that the field is small today and fullfills nucleosynthesis constraints.
We will therefore now discuss the cosmology of the radion field with a potential energy of
the form (\ref{poti}) and therefore get an estimate of $\phi_\infty$, which determines
the coupling to matter locally (see eq. (\ref{radioncoupling})).

\section{Quintessential Potentials for the Radion}
In this section, we will switch notation and will denote the radion field as $\phi$.
We will discuss the different cosmological epochs separately. Before imposing 
the local constraints, however, we will discuss the cosmological evolution in 
general.

\subsection{The inflationary era}
Let us consider first an inflationary era in order to find the initial conditions for the
radion field in the radiation dominated epoch. For the inflationary model, we choose a
chaotic inflationary potential of the form
\begin{equation}
V(\sigma) = \frac{1}{2}m^2 A^2 \sigma^2.
\end{equation}
where $\sigma$ has dimension one.
We assume that the inflaton field $\sigma$ is confined on the
positive tension brane. Then, in the Einstein frame, its mass
depends on the moduli fields through a factor of $A$. The
Klein--Gordon equation for the field $\phi$ reads
\begin{equation}
\ddot\phi +3H\dot \phi + \kappa_4^2 \frac{\partial V(\phi,\sigma )}{\partial
  \phi}=-\kappa_4^2T\alpha_\phi
\end{equation}
where $T=-\dot \sigma^2 -2m^2 A^2\sigma^2$ is the trace of the
energy momentum tensor of $\sigma$. Thus, the dynamics of the
field $\phi$ are governed by an effective potential
\begin{equation}
V_{\rm eff}=\frac{\Lambda^4}{\phi^\alpha}+\frac{3}{2}(1+\frac{\beta}{2}\phi^2)
m^2 \sigma^2,
\end{equation}
where we have neglected the kinetic energy of $\sigma$ as the field is assumed to
slow--roll. During inflation, $\sigma = \sigma_{\rm inf}$ is almost constant.
The minimum of the effective potential is given by
\begin{equation}
\phi_0= \left(2\frac{\alpha\Lambda^4}{3\beta   m^2
 \sigma_{inf}^2}\right)^{1/(\alpha +2)}
\end{equation}
and the  effective mass of the $\phi$ field at the minimum is
\begin{equation}
m^2_{\rm eff}=3\beta (\alpha +2 ) H^2,
\end{equation}
where $H^2 = m^2 \sigma^2_{\rm inf}/2m_{\rm Pl}^2$. This large mass guarantees
that the effective potential for $\phi$ is very steep around its
minimum and $\phi$ decays rapidly towards the minimum.
Thus, just before the end of inflation, $\phi$ sits at the minimum 
of the effective potential. Notice that the value of $\phi$ is tiny, 
as $\Lambda$ is small. We will assume that the field $\phi$ is a fundamental 
field and therefore that the inflaton field does only decay into radiation, 
the standard model particles and dark matter. Thus, during reheating no 
energy will be released into the field $\phi$ and therefore during 
reheating the field remains small. This provides a hint about the 
initial value of the radion at the beginning of the radiation epoch.

\subsection{The radiation-- and matter--dominated epoch}

Let us consider now the cases of the radiation and matter
dominated epochs. The Klein--Gordon for the radion equation reads
\begin{equation}
\ddot\phi +3H\dot \phi + \kappa_4^2 \frac{\partial V(\phi)}{\partial
  \phi} = - \kappa_4^2 \rho_m \phi,
\end{equation}
that is, the effective potential reads
\begin{equation}
V_{\rm eff}=\frac{\Lambda^4}{\phi^\alpha}+{\rho_m}
\left(1+\frac{\beta}{2}\phi^2 \right).
\end{equation}
The field value at the minimum is given by
\begin{equation}
\phi_{\rm min}= \left(\frac{\alpha\Lambda^4}{\beta
\rho_m}\right)^{1/(\alpha +2)}.
\end{equation}
Because the field is small, we have that
\begin{equation}
\rho_m\approx\frac{\rho_m^0}{a^3}.
\end{equation}
Now we can consider two cases during radiation domination:
either the fields starts at $\phi(t_i) > \phi_{\rm min}$ or
at $\phi(t_i)< \phi_{\rm min}$. Assume that the field value
is larger than the minimum. Then we can neglect the potential
and the field is governed by
\begin{equation}
\ddot \phi +3H\dot \phi \approx -\frac{\rho_m}{m^2_{pl}} \phi.
\end{equation}
This is the equation of an damped oscillator. Because the mass is
much smaller than the Hubble parameter, the field is overdamped
and does not move. Only if $H$ drops below the mass the field
starts to evolve and moves towards the minimum. Let us consider
now the case that $\phi<\phi_{\rm min}$. If the potential energy
is initially very large, the field rolls down and becomes
kinetically dominated ($\Omega_i$ is the initial fraction of energy
stored in the radion field) (see also \cite{tracker})
\begin{equation}
\phi= \sqrt{6\Omega_i}(1-\frac{a_{\rm ini}}{a})
\end{equation}
with an equation of state $w=1$. It stops at
\begin{equation}
\phi_{\rm stop} = \sqrt{6\Omega_i}
\end{equation}
with an equation of state $w=-1$.  This can be seen in figure 3
where a plateau is reached before converging to the attractor (in
Planck units)
\begin{equation}\label{kgsolution1}
\phi_{\rm rad}(t) = \left(\frac{2\alpha \Lambda^{4}}{(2\gamma^2 + \gamma)m_{ pl}^4}\right)^{1/(2+\alpha)}
(m_{pl}t)^{\gamma},
\end{equation}
with $\gamma = 2/(2+\alpha)$. The equation of state is given by
\begin{equation}\label{eosradiation}
w=\frac{(\alpha-2)\gamma - 1}{(\alpha+2)\gamma +1}
\end{equation}
corresponding to the usual Ratra--Peebles attractor \cite{rpattractor}. 

In the matter dominated era, the minimum of the effective potential
is located at
\begin{equation}
\phi_{\rm min}^{\alpha +2}= \frac{\alpha \Lambda^4}{\rho_m}
\end{equation}
The mass at the minimum ($V''$) is then
\begin{equation}
m^2= (\alpha+2) \rho_m
\end{equation}
in Planck units. The Klein--Gordon equation has an attractor
\begin{equation}\label{kgsolution2}
\phi_{\rm matt}(t) = \left(\frac{\alpha \Lambda^4}{m_{pl}^4(\gamma^2
+\frac{\gamma}{2} +\frac{4}{3}\beta })\right)^{\frac{1}{\alpha
+2}}(m_{pl}t)^\gamma
\end{equation}
 implying that
\begin{equation}
\frac{\phi_{\rm matt}}{\phi_{\rm min}}=\left(\frac{4}{4\beta +3\gamma^2
+3\frac{\gamma}{2}}\right)^{\frac{1}{\alpha +2}}
\end{equation}
Notice that $\phi_{\rm matt} <\phi_{\rm min}$. The field follows
the minimum but does not settle at the minimum (see figure 3). The
equation of state reads
\begin{equation} \label{eosmatter}
w=\frac{(\alpha-2)\gamma^2 - \gamma - \frac{8}{3}\beta}
{(\alpha+2)\gamma^2 + \gamma + \frac{8}{3} \beta},
\end{equation}
which is smaller than in the radiation dominated era.

 We have numerically solved the equations
and compared it to the ordinary quintessence model in General
Relativity. The results are shown in Figures 1, 2 and 3.

\begin{figure}[!ht]
\begin{center}
\epsfig{file=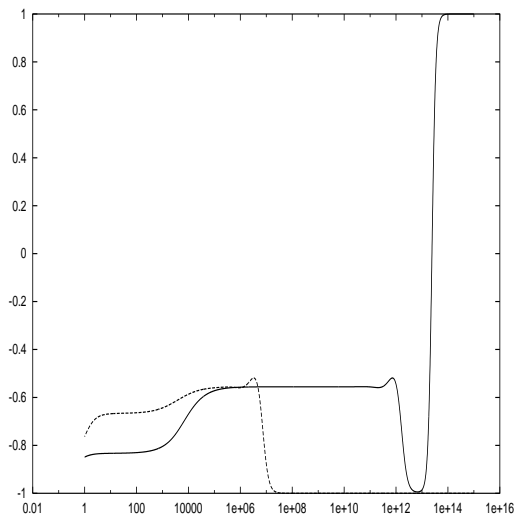,height=7cm,width=7cm,angle=270}
\end{center}
\caption{Evolution of the equation of state for the model in
General Relativity (dashed line) and in the brane model (solid
line) as a function of $1+z$. It can be seen that the equation of
state today is nearer $-1$ in the brane model than in General
Relativity. The exponent in the inverse power--law potential is
$\alpha=1$, for which $\gamma=2/3$.}
\end{figure}
\begin{figure}[!ht]
\begin{center}
\epsfig{file=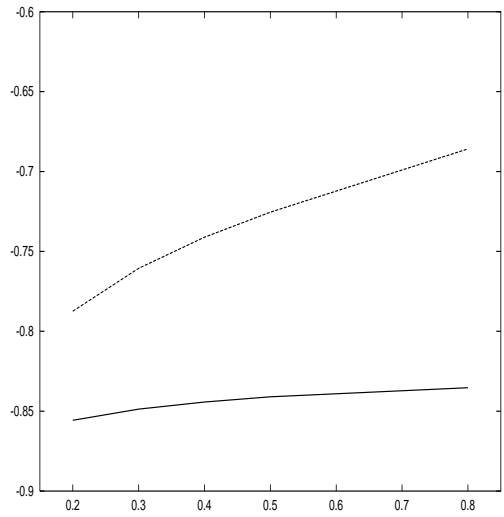,height=7cm,width=7cm,angle=270}
\end{center}
\caption{Dependence of the equation of state as a function of
$\Omega_{\rm m}$. As can be seen, the dependence is weaker in the
brane model (solid line) than in General Relativity (dashed line).
For this plot, the exponent in the inverse power--law potential is
$\alpha=1$.}
\end{figure}

In Figure 1 we plot the evolution of the equation of state as a
function of $1+z$ for the case of $\alpha=1$ and the same initial
conditions. We compare the brane theory together with a
quintessence model in General Relativity. As  can be seen, the
evolution of the equation of state around $z=0$ is more pronounced
in General Relativity than in the brane model. Furthermore, the
equation of state is smaller in the brane scenario than in General
Relativity.

\begin{figure}[!ht]
\begin{center}
\epsfig{file=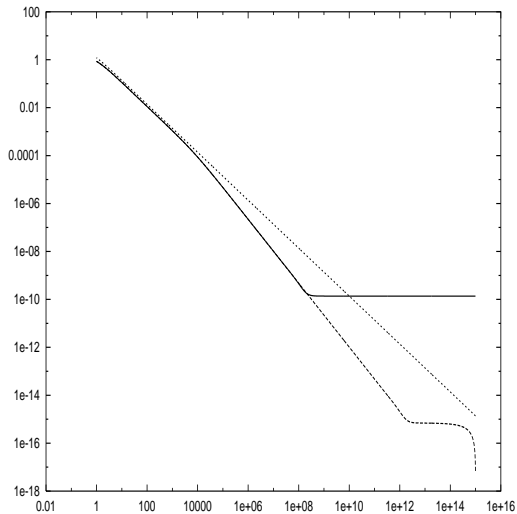,height=7cm,width=7cm,angle=270}
\end{center}
\caption{Evolution of the field as a function of $1+z$ for
different initial conditions. The straight dashed line shows the
position of the minimum of the effective potential.}
\end{figure}

In Figure 2 we show the dependence of the equation of state $w$
as a function of $\Omega_{\rm m}$. It can be seen that in General
Relativity this dependence is stronger than in the brane model. Finally,
in Figure 3 we plot the evolution of the radion field (in Planck units)
as a function of $1+z$ for different initial conditions. As  can be seen,
the field evolves according to the solution (\ref{kgsolution1}) in the
radiation dominated epoch and according to (\ref{kgsolution2}) in the
matter dominated epoch.

\subsection{Imposing local constraints}
The discussion on the cosmological evolution so far did not take
into account the constraints imposed by local experiments. In
fact, if the radion field acts as dark energy, its evolution is
similar to a field with inverse power--law potential in General
Relativity. In particular, for $\alpha={\cal O}(1)$, the field
value today is of order the Planck mass. This, however, is not 
compatible with local experiments. Imposing that $\theta \le 
10^{-5}$ leads to $\Lambda \approx 10^{-3}$ eV and $\alpha \leq 
10^{-5}$. The potential is extremely flat and peaked sharply 
around $\phi=0$. Thus, for the potential considered here the price 
to be paid is that, at least for the standard potential used here, 
the parameters have to be extremely fine--tuned, if the radion should 
play the role of dark energy. 

\section{The case for a non--standard potential}
So far we have focused on a particular inverse power law potential, in 
which $V(\phi) \rightarrow 0$ for $\phi \rightarrow \infty$.
In the following we will consider another class of potentials leading to 
an interesting cosmology with no conflict with local experiments. The 
potential is not a quintessential potential, however, but has the form
\begin{equation}\label{champot}
V(\phi)= M^4 e^{\left(\frac{M}{m_{pl}\phi}\right)^n}
\end{equation}
where $M=10^{-3}$ eV in order to trigger acceleration now. It is not a
quintessence potential per se as it does not converge to zero at
infinity. It should be considered as a very flat potential which may 
appear from some non-perturbative physics for the radion. As we will now 
see, such a potential leads to acceleration and drives the field towards
very small values ($\phi_\infty \ll 1$).

For the potential above, the minimum of the effective potential is given by
\begin{equation}
\phi_{min}^{n+2}=n\left(\frac{M}{m_{pl}}\right)^{n} \frac{V(\phi_{min})}{\beta \rho_m}.
\end{equation}
For $V=O( M^4)$ this leads to
\begin{equation}
\phi_{\infty}^{n+2}=O\left(\frac{M}{m_{pl}}\right)^{n}.
\end{equation}
Notice that $\phi_{\infty}=O(10^{-10})$ for $n=1$ implying that solar tests are
automatically satisfied.

Let us discuss the cosmology of this model.
The mass of the radion at the minimum reads
\begin{equation}
\frac{m^2}{H^2}= 3\beta \Omega_m \left[ n(n+2)  + n^2 (\frac{M}{m_{pl}})^n
  \frac{1}{\phi^n} \right].
\end{equation}
There are two regimes. When $m_{pl} \phi \ll M$, the mass is very large
$M\gg H$ implying that the minimum acts as an attractor. When 
$m_{pl}\phi \gg M$, the potential reads
\begin{equation}
V(\phi) =M^4 + \frac{\Lambda ^4}{\phi^n},
\end{equation}
where
\begin{equation}
\Lambda^4= \frac{M^{4+n}}{m_{pl}^n}.
\end{equation}
Now as long as $V(\phi)$ is subdominant, the constant term is
irrelevant and we retrieve the analysis of the power law
potential. This implies that in the radiation and matter dominated
eras, the radion follows the attractors that we found previously.
Only the early Universe is  modified. 

We have seen that  $\phi_{now} \ll 1$ and therefore  solar tests are
 satisfied.
 What about laboratory
tests of the equivalence principle? Let us consider the gravity
 experiments  in a cavity of radius $R_*$. Inside
the vacuum chamber the field is early constant and determined by
\begin{equation}
V''\vert_{\phi_{cav}}= m^2_p R_*^{-2},
\end{equation}
i.e. the Compton wavelength of the radion is then given by $R_*$.
Moreover the field in the cavity is larger than the field outside the
cavity implying that
\begin{equation}
\vert \frac{\partial \phi}{\partial r}\vert \le \frac{\phi_{cav}}{R_*}.
\end{equation}
Inside the cavity, a gradient of $\phi$ leads to a new force acting on
particles
and a  breaking of  the equivalence principle to the accuracy level
\begin{equation}
\eta= 10^{-4}  \beta \frac{\phi_{cav}^2}{R_*}\frac{4\pi m_{pl}^2
  R_{earth}^2}{M_{earth}}
\end{equation}
Imposing that $\eta \le 10^{-26}$ \cite{st-test2} leads to
\begin{equation}
\phi_{cav}\le 10^{-13}.
\end{equation}
which is always satisfied.

To conclude this Section, for a potential of the form (\ref{champot})  
the mass scale has to be chosen such that $M \approx 10^{-3}$~eV if  
the radion is to be a dark energy candidate. This tuning is no more
than that required by a cosmological constant.

\section{Conclusions}
In this paper, the chameleon mechanism has been investigated in the 
case of field dependent couplings. Such couplings arise naturally in 
theories with extra dimensions, such as brane world scenarios. In 
theories with constant couplings a thin--shell regime exists, in which 
the deviations from General Relativity are suppressed. We 
have shown, that this is not true in models with 
field--dependent couplings. In these models the 
thin--shell regime does not exist (i.e. no suppression 
of the couplings) and, hence, there are potentially large 
deviations from General Relativity. 

Using the low-energy effective action, we have 
considered the cosmology of the radion field at 
large distances on 
the branes (or low energy). Possible deviations from General Relativity due
to the smallness of the radion mass were investigated. Two regimes can be distinguished, 
leading to the thin and thick shell properties. At short
distances between the branes, the radion satisfies the thin-shell 
property and no
deviation can be detected either on earth or in solar system
experiments. Satellites experiments would be able to 
detect large deviations from gravity.
For large distances between the branes, the thin-shell property is replaced by a
thick--shell regime. In this case deviations from normal gravity are
small provided that the cosmological value of the radion is small 
enough. We have investigated a class of potentials for which the
radion (at large distances between the branes) 
can both act as quintessence and satisfies gravity tests. 
Unfortunately, future satellite experiments would not notice the presence 
of the radion as the coupling in that case is too small. 

If the branes are close together, the model and its cosmology 
are similar to the original chameleon model \cite{chameleon,chameleoncosmology}, 
local constraints are satisfied and the radion can also act as a dark energy 
candidate. In this case, future satellite experiments could detect 
the predicted deviations from General Relativity.  

\vspace{0.5cm}

\noindent{\bf Acknowledgements:} We are grateful to Justin Khoury for discussions. 
We wish to thank the British Council--Alliance for an exchange grant. CvdB and ACD 
are supported in part by PPARC.

\section{Referenes}

\end{document}